\documentclass[prd,aps,showpacs,,nofootinbib,twocolumn,superscriptaddress,preprintnumbers,epsf,psf]{revtex4}
\usepackage[english]{babel}
\usepackage{pstricks}
 \usepackage{epsfig}
\usepackage{epsf}
\usepackage{amsmath}
\usepackage{amsfonts}
\usepackage{amssymb}
\usepackage{slashed}

\begin{document}

\title{Space-time structure of polynomiality and
positivity for GPDs}
\author{ I.~V.~Anikin}
\email{anikin@theor.jinr.ru}
\affiliation{Bogoliubov Laboratory of Theoretical Physics, JINR,
             141980 Dubna, Russia}
\author{ I.~O.~Cherednikov}
\email{igor.cherednikov@uantwerpen.be}
\affiliation{Departement Fysica, Universiteit Antwerpen, B-2000 Antwerpen, Belgium}
\affiliation{Bogoliubov Laboratory of Theoretical Physics, JINR,
             141980 Dubna, Russia}
\begin{abstract}
We study the space-time structure of polynomiality and positivity---the most important properties which are inherent to
the generalized parton distributions (GPDs).
In this connection, we re-examine the issue of the time- and normal- ordering in the operator definition of GPDs.
We demonstrate that the contribution of the anti-commutator matrix element in the collinear kinematics, which was previously argued to vanish, has to be added in order to satisfy the polynomiality condition.
Furthermore, we schematically show that a new contribution due to the anti-commutator modifies likewise the so-called positivity constraint, i.e., the Cauchy-Bunyakovsky-Schwarz inequality, which is another important feature of the GPDs.
\end{abstract}
\pacs{12.38.Bx, 13.60.Le}
\date{\today}
\maketitle

\section{Introduction}

The space-time structure of the generalized parton distributions (GPDs), together with their polynomiality, is encoded in the matrix elements of the (anti)commutators of the fermion fields.
In this connection, the problem of the time-ordering and the consistency of the replacement of it by the ordinary ordering in the generalized parton distributions (GPDs) is discussed in the literature since many years (see, e.g., \cite{Jaffe, Diehl}). In the cases of the DIS and DVCS processes, it was argued the matrix element of the fermion anti-commutator
vanishes and, therefore, the time-ordering in GPDs is ``illusory'' and it can readily be replaced by the ordinary ordering of the corresponding fermion operators. The crucial point of those studies was that the anti-commutator contribution is defined by the limit of $1/(k^-)^{n-1}$ where $n\geq 2$ at $k^-\to\infty$ for the Mandelstam variables differ from zero.
Furthermore, it was shown in Ref. \cite{Diehl} that the support and spectral properties of the GPDs emerge naturally.

The purpose of this paper is to demonstrate that in the collinear kinematics and within the
factorization procedure in the $t$-channel, where the Mandelstam variable $t$ is small compared to $s$,
the matrix element of the fermion anti-commutator does not vanish and
yields a term necessary to hold the model-independent polynomiality condition for any kind of the generalized parton distributions.
Moreover, the latter is even valid in the regime where the Mandelstam variables $s$, $u$ and $t$ are similarly small, that is in the so-called totally
collinear kinematics. Note that this particular point $s\sim t\sim 0$ in the Mandelstam plane is responsible for the duality regime of the factorization, discussed in detail in Ref. \cite{ACST-duality}, and bridges between the factorizations in the $t$- and $s$-channels. The comprehensive analysis of this very interesting point is forthcoming in
\cite{ACT-Positiv}.
We also demonstrate schematically that the obtained contribution, arising from the matrix element of the fermion anti-commutator, modifies evenly another important property of the GPDs, the positivity. We show, moreover, that this modification allows us to relate the GPDs with the non-perturbative fermion condensates.

\section{Heisenberg and Interaction representations}

As the first step, let us start with the outline of the main issues of the matching between the Heisenberg and interaction
representations. Consider, for instance, the time-ordered product of two fermion fields in the interaction representation with the $\mathbb{S}$-matrix,
$\mathbb{S}(t_2=\infty,\, t_1=-\infty)\equiv\mathbb{S}_{\infty,-\infty}\equiv\mathbb{S}$. Using the Wick theorem,
\begin{eqnarray}
\label{IntReps1}
&&\mathrm{T}
\psi(x) \, \bar\psi(y)\, \mathbb{S}_{\infty,-\infty}
= G^c(x,y) +
\nonumber\\
&&\sum\limits_{n}\frac{(ig)^n}{n!} \int (d^4\xi)_n
{\sum\limits_{\rm{pairing}}}^\prime
 : \psi(x)\, \bar\psi(y)\, (\bar\psi \hat A \psi)_{\xi_1}
 \nonumber\\
&&\ldots
 (\bar\psi \hat A \psi)_{\xi_n} :\, ,
\end{eqnarray}
where $:...:$ denotes the normal-ordered product of fields. Here ${\sum\limits_{\rm{pairing}}}^\prime$ stands for the sum of all possible sets of contractions (or pairings) between the fields
excluding the terms with {\it all} fields contracted, the latter being accumulated in $G^c(x,y)$.

The field $\psi (x)$ in the interaction representation
transforms into the Heisenberg field operator $\psi_H(x)$ as follows
$\psi_H(x)=\mathbb{S}^\dag_{t,0}\psi(x)\mathbb{S}_{t,0}$.
By making use of this transformation, we obtain the relation between the time-ordered products of two fermion fields in the Heisenberg and in the interaction representations, respectively:
\begin{eqnarray}
\label{TprodInH}
\mathrm{T}
\psi(x) \, \bar\psi(y)\, \mathbb{S}_{\infty,-\infty}
=
\mathbb{S}_{\infty,0}\,
\mathrm{T}
\psi_H(x) \, \bar\psi_H(y)
\, \mathbb{S}_{0,-\infty}\, .
\end{eqnarray}
Calculating the vacuum expectation value of the time-ordered operator product, we get the standard definition of the connected Green function:
\begin{eqnarray}
\label{GrF1}
S^c(x,y)&=& \frac{\langle 0| \mathrm{T}
\psi(x) \, \bar\psi(y)\, \mathbb{S}_{\infty,-\infty}|0 \rangle}{\mathbb{S}_0}
\nonumber\\
&=&{~}^H\langle 0| \mathrm{T}
\psi_H(x) \, \bar\psi_H(y)|0 \rangle^H,
\end{eqnarray}
where the normalization condition $\mathbb{S}_0=\langle 0| \mathbb{S}_{\infty,-\infty}|0 \rangle$ cancels all contributions from the disconnected graphs in the interaction representation, while the vacuum state in the Heisenberg picture is defined as ${~}^H\langle 0|= \langle 0| \mathbb{S}_{\infty,0}$ and
$|0 \rangle^H =  \mathbb{S}_{0,-\infty}  |0 \rangle$. In what follows we shall only keep the up-script $H$ in formulae
to indicate the Heisenberg representation.

If we consider now the hadronic matrix element of the time-ordered operator product instead of
the vacuum average, we observe (upon application of the Wick theorem) that the terms related to the matrix elements of the normal-ordered operators {\it do not disappear}. Notice that the same inference is true if our states are the physical or
non-perturbative vacuum. At the same time, the fully-contracted terms refer to the disconnected matrix elements
and, therefore, have to be discarded. Indeed, we have
\begin{eqnarray}
\label{HadMeS}
&&\langle p_2| \mathrm{T}
\psi(x) \, \bar\psi(y)\, \mathbb{S}_{\infty,-\infty}|p_1 \rangle =
G^c(x,y)\langle p_2|p_1 \rangle +
\\
&&\sum_{n;i,j} \int (d^4\xi)_n \langle p_2|
: \psi(\xi_i) C_n(\xi_i,\xi_j; x,y) \bar\psi(\xi_j) :|p_1 \rangle +
\nonumber\\
&&(``N>2 \,\, \hbox{normal-ordered \,\,operators}") \, ,
\nonumber
\end{eqnarray}
where $C_n(\xi_i,\xi_j; x,y)$ is the corresponding product of different propagators. The first term in the {\it l.h.s.} of (\ref{HadMeS}), $G^c(x,y)\langle p_2|p_1 \rangle$,
which is proportional to $\delta^{(4)}(p_2-p_1)$, describes only the disconnected Feynman diagrams.
Thus, we define the connected matrix element of the time-ordered operator product as
\begin{eqnarray}
\label{HadMeSCon}
&&\langle p_2| \mathrm{T}
\psi(x) \, \bar\psi(y)\, \mathbb{S}_{\infty,-\infty}
|p_1 \rangle_C =
\nonumber\\
&&\sum_{n;i,j} \int (d^4\xi)_n \langle p_2|
: \psi(\xi_i) C_n(\xi_i,\xi_j; x,y) \bar\psi(\xi_j) :|p_1 \rangle\, +\,
\nonumber\\
&&(``N>2 \,\, \hbox{normal-ordered \,\,operators}") \, ,
\end{eqnarray}
where the subscript $C$ points out that we are dealing with the connected matrix elements.
On the other hand, the hadron matrix element (\ref{HadMeSCon}) can be written in compact
form in the Heisenberg representation. We have
\begin{eqnarray}
\label{HadMEHeisCon}
&&\sum_{n;i,j} \int (d^4\xi)_n \langle p_2|
: \psi(\xi_i) C_n(\xi_i,\xi_j; x,y) \bar\psi(\xi_j) :|p_1 \rangle\, +
\nonumber\\
&&(``N>2 \,\, \hbox{normal-ordered \,\,operators}") \equiv
\nonumber\\
&&\langle p_2|
:\psi(x) \, \bar\psi(y): |p_1 \rangle^H_C \, ,
\end{eqnarray}
or, comparing Eq. (\ref{HadMeSCon}) with Eq. (\ref{HadMEHeisCon}), we conclude that
\begin{eqnarray}
\label{MatHaI}
\langle p_2| \mathrm{T}
\psi(x)\bar\psi(y)\mathbb{S}
|p_1 \rangle_C =
\langle p_2|
:\psi(x)\bar\psi(y): |p_1 \rangle^H_C\, .
\end{eqnarray}
In turn, given that we consider only the connected matrix elements, the normal-ordered operators in the Heisenberg representation can be replaced by the time-ordered operators
\begin{eqnarray}
\label{MatHaI2}
\langle p_2|
:\psi(x) \, \bar\psi(y): |p_1 \rangle^H_C=
\langle p_2|
\mathrm{T}\psi(x) \, \bar\psi(y)|p_1 \rangle^H_C\, .
\end{eqnarray}
Let us emphasize that Eqs. (\ref{MatHaI}) and (\ref{MatHaI2}) are our principal observations, to which we would like to attract attention of the reader.

\section{The factorized DVCS amplitude}

Now we concentrate on the DVCS amplitude factorized into the hard and the soft parts. Before the factorization is carried out, the DVCS amplitude in the interaction picture can be expressed as
\begin{eqnarray}
\label{DVCS1}
{\cal A}_{\mu\nu}=e^2\hspace{-0.2cm}\int\hspace{-0.1cm} d\xi d\eta
e^{-iq\cdot\xi + iq^\prime\cdot\eta}
\langle p_2| \mathrm{T}
J^{em}_\nu(\eta)J^{em}_\mu(\xi)\mathbb{S}|p_1 \rangle_C\, ,
\nonumber
\end{eqnarray}
where $J^{em}_\mu$ is the electromagnetic current and the $\mathbb{S}$-matrix involve all possible interactions. Expanding the $\mathbb{S}$-matrix in power of the coupling constant (we do not need yet to specify the Lagrangians we are working with) and making use of the Wick theorem, we obtain the standard expression for the amplitude
\begin{eqnarray}
\label{AmpDVCS}
{\cal A}\Rightarrow
\langle p_2| :\bar\psi(\eta)
\underline{\gamma_\nu\,S(\eta-\xi)\,\gamma_\mu}\psi(\xi): |p_1 \rangle_C +\ldots \,,
\nonumber
\end{eqnarray}
where the ellipsis denotes other possible combinations of the normal-ordered operators including
the cross-terms. We here underlined the combination to stress that it will form the hard part of the
amplitude. Notice that the combinations with $N>2$ normal-ordered operators are not the issues in the present paper.

The factorization of the amplitude  in the interaction representation consists in the separation the hard part (underlined) from the soft part (which will be expressed in what follows in terms of the GPDs):
\begin{eqnarray}
\label{GPDs1}
&&\Phi(x,\xi)= \int d^4k \, \delta(x-k \cdot n)\, d^4z \ e^{i(k-\Delta/2) \cdot z} \times
\nonumber\\
&&\langle p_2| \mathrm{{\tilde T}}
\bar\psi(0) \psi(z) \mathbb{S}_{\infty,-\infty}|p_1 \rangle_C,
\end{eqnarray}
where $\mathrm{{\tilde T}}$ suggests that we have to hold only two fermion operators as the normal-ordered one.
The spinors should be understood as the operators with the free Dirac indices.
As it has been mentioned above, the Heisenberg representation
allows us to re-write the {\it r.h.s.} of Eq. (\ref{GPDs1}) in the most compact form as
\begin{eqnarray}
\label{GPDs2}
&&\Phi(x,\xi)= \int d^4k \, \delta(x-k \cdot n)\, d^4z \ e^{i(k-\Delta/2) \cdot z} \times
\nonumber\\
&&
\langle p_2|
:\bar\psi(0) \psi(z): |p_1 \rangle^H_C \, .
\end{eqnarray}
Given that we are again interested in the connected matrix elements only, we may write the time-ordered operators instead of the normal-ordered operators in the Heisenberg representation,
{\it i.e.}
\begin{eqnarray}
\label{GPDs3}
&&\Phi(x,\xi)= \int d^4k \, \delta(x-k \cdot n)\, d^4z \  e^{i(k-\Delta/2) \cdot z} \times
\nonumber\\
&&
\langle p_2|
\mathrm{T}\bar\psi(0) \psi(z)|p_1 \rangle^H_C \, .
\end{eqnarray}
Alternatively, using the light-cone notations, one has
\begin{eqnarray}
\label{GPDs4}
\Phi(x,\xi)
=\int dk^- d^2{\bf k}_T \Phi(xP^+,k^-, {\bf k}_T; \xi) .
\end{eqnarray}
These three representations, Eqs. (\ref{GPDs2})-(\ref{GPDs4}), are equivalent. Recall that the function $\Phi$ possesses the free Dirac indices. If we now project the GPDs (\ref{GPDs2})-(\ref{GPDs4}) to the $\gamma^+$-matrix, we shall obtain the various twist-$2$ generalized parton distributions, depending on the hadron target:
\begin{eqnarray}
\label{Hfun}
\Phi^{[\gamma^+]}\stackrel{{\rm def}}{=}{\rm tr}[ \gamma^+\,\Phi]
\Rightarrow \{ H_1; H, E; ... \}.
\end{eqnarray}
We can thus conclude that since we deal only with the connected matrix elements, the time-ordering and/or the normal-ordering occur in the GPDs of any kind in an equivalent way. This is one of our main observations.

Let us now focus on Eq. (\ref{GPDs3}). It is well-known that the time-ordered combination of spinors can be expressed through their commutator and anti-commutator:
\begin{eqnarray}
\label{GPDs5}
\Phi(x)=\Phi^{[...]}(x) +  \Phi^{\{...\}}(x) ,
\end{eqnarray}
where
\begin{eqnarray}
\label{PhiCom}
&&\Phi^{[...]}(x)=\frac{1}{2}
\int d^4k \, \delta(x-k.n)\, d^4z e^{i(k-\Delta/2).z}\times
\nonumber\\
&&\langle p_2|
[\bar\psi(0), \psi(z)]|p_1 \rangle^H_C\, ,
\end{eqnarray}
and
\begin{eqnarray}
\label{PhiAntiCom}
&&\Phi^{\{...\}}(x)=\frac{1}{2}
\int d^4k \, \delta(x-k.n)\, d^4z e^{i(k-\Delta/2).z} \times
\nonumber\\
&&
\varepsilon(z_0) \,
\langle p_2| \{\bar\psi(0), \psi(z)\}|p_1\rangle^H_C .
\end{eqnarray}
We would like to emphasize that the presence of $\varepsilon(z_0)$ in Eq. (\ref{PhiAntiCom}) leads to the
absence of any $s (u)$-channel cuts in the anti-commutator contribution, while the commutator contribution
can be related to the $s (u)$-channel cuts.
Indeed, consider the first term of the anti-commutator contribution, see Eq. (\ref{PhiAntiCom}):
\begin{eqnarray}
\frac{1}{2}
\int\limits_{-\infty}^{\infty}d^4z \, \varepsilon(z_0)
e^{i(k-\Delta/2).z}
\langle p_2|\bar\psi(0) \psi(z)|p_1\rangle^H_C.
\end{eqnarray}
Inserting of the full set and making use of the translation invariance, one presents this expression in the following form:
\begin{eqnarray}
&&\sum_X\hspace{-0.5cm}\int \frac{i}{\pi}{\cal P}\frac{1}{k_0-P_0+P_0^X}
\delta^{(3)}(\vec{k}-\vec{P}+\vec{P_X})\times
\nonumber\\
&&\langle p_2|\bar\psi(0)|P_X\rangle^H_C
\langle P_X|\psi(0)|p_1\rangle^H_C.
\end{eqnarray}
One can see that the four-dimensional $\delta$-function, needed for
the appearance of the cut in $s(u)$-channel, is absent. The similar is valid for the
second term of Eq. (\ref{PhiAntiCom}).

It is obvious that if the anti-commutator were vanishing for some reason (see, e.g., \cite{Jaffe, Diehl}),
it would be permitted to replace the time-ordering by the ordinary product of operators.
That is to say, the time-ordering gets ``illusory''.

However, we here present an alternative approach to show that the contribution of the anti-commutator, $\Phi^{\{...\}}(x)$, does not vanish
in the case of factorization in the $t$-channel, using
the collinear kinematics (see below), where the Mandelstam variable $t$ is small compared to $s$.
One of our main evidences is that the contribution of the anti-commutator matrix element
is necessary to obey the model-independent {\it polynomiality condition} for the GPDs, which arises from the requirement of the Lorentz covariance of
the corresponding matrix element. We will demonstrate this by taking as an example the box diagram within a toy model which was very useful to the introduction of GPDs \cite{Rad97}.

\section{A toy model for the box diagram}

Consider first the box diagram contribution to the DVCS amplitude:
\begin{eqnarray}
\label{Process}
\gamma^*(q) + A(p_1) \to \gamma(q^\prime) + A(p_2)
\end{eqnarray}
in perturbation theory.
The box diagram is the most illustrative object to reveal the main features of the factorization approach involving
the GPDs, see \cite{Rad97}. Because the factorization procedure is extensively described in the literature,
we will skip the details of this procedure. We begin with the definition of the
light-cone kinematics, which we will use in what follows:
\begin{eqnarray}
\label{kin}
&&n^2=p^2=0,\quad p\cdot n=1,
\nonumber\\
&&
g_{\mu\nu}^T=g_{\mu\nu}-p_\mu n_\nu-p_\nu n_\mu,
\nonumber\\
&&
p_2=(1-\xi) p +(1+\xi)\frac{\bar M^2}{2} n  + \Delta_T/2\,,
\nonumber\\
&&
p_1=(1+\xi) p +(1-\xi)\frac{\bar M^2}{2} n  - \Delta_T/2\,,
\nonumber\\
&&
q^\prime= P.q^\prime n ,
\nonumber\\
&&
\bar Q=(q+q^\prime)/2,
\quad
P=(p_1+p_2)/2,
\quad
\Delta=p_2-p_1,
\nonumber\\
&& P^2=\bar M^2=\frac{\Delta_T^2-t}{4\xi^2}, \quad \Delta^2=t.
\end{eqnarray}
Without the loss of generality, we may use the {\it collinear} kinematics which corresponds to the case when $\Delta_T\approx 0$.
\begin{widetext}
\begin{figure}[t]
 $$\includegraphics[angle=90,scale=0.75]{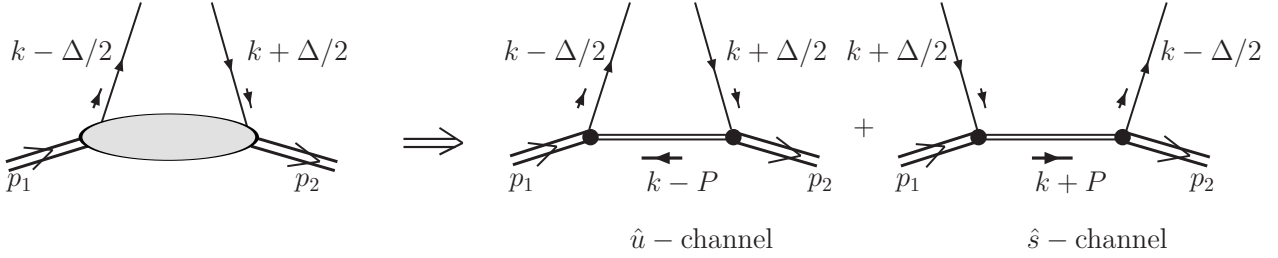}$$
   \caption{\label{fig:1}GPDs within a toy model.}
\end{figure}
\end{widetext}
We now approach the factorized amplitude in perturbation theory, so that we can write in the twist-$2$ level:
\begin{eqnarray}
\label{Boxdia}
&&{\cal A}_{\mu\nu}=\int\limits_{-1}^{1} dx \,
{\rm tr}[ \gamma_\nu S(xP+\bar Q)\gamma_\mu \gamma^-]\times
\nonumber\\
&&\int d^4k\, \delta(x-k\cdot n)\, \Phi^{[\gamma^+]}(k) + \hbox{``crossed''}\, .
\end{eqnarray}
We identify the initial and final states in the corresponding matrix elements with the electron/quark states. In this case, the soft part of this amplitude takes the following form
(in the Feynman gauge), see Fig.1:
\begin{eqnarray}
\label{PhiPT}
&&\Phi^{[\gamma^+]}(x,\xi)=\int (d^4k)\, \delta(x-k\cdot n)\, \Phi^{[\gamma^+]}(k)
\stackrel{g^2}{=}
\nonumber\\
&& ig^2 \int (d^4k)\, \delta(x-k\cdot n)\, D(k-P)\times
\\
&&[ \bar u(p_2)\, \gamma_\alpha \,S(k+\Delta/2)\, \gamma^+ \,S(k-\Delta/2)\, \gamma_\alpha\, u(p_1)] \, .
\nonumber
\end{eqnarray}

Making use of Eq. (\ref{kin}), we obtain that
\begin{eqnarray}
\label{Denom}
& &(k-\Delta/2)^2=2k^- p^+ (x+\xi) - (x+\xi)\xi \bar M^2 - {\bf k}_T^2 \ ,
\nonumber\\
& &(k+\Delta/2)^2=2k^- p^+ (x-\xi) + (x+\xi)\xi \bar M^2 - {\bf k}_T^2 \ .
\nonumber\\
\end{eqnarray}
For the parton subprocess, we also introduce the corresponding Mandelstam variables:
\begin{eqnarray}
\label{MandSubProc}
&&\hat s= (k+P)^2=2k^- p^+ (x+1) + (x+1)\bar M^2 - {\bf k}_T^2,
\nonumber\\
&&\hat  u=(k-P)^2=2k^- p^+ (x-1) + (1-x)\bar M^2 - {\bf k}_T^2.
\nonumber\\
\end{eqnarray}
Notice that within the collinear kinematics, $\bar M \approx \sqrt{-t}/(2\xi)$, and,
therefore, it can be discarded with respect to the large $p^+$. At the same time,
keeping the terms which are proportional to $t$ will never
allow the poles to jump from the upper plane to the lower one.

For the sake of simplicity, we extract the following structure integral:
\begin{eqnarray}
\label{AmpStInt}
\Phi^{[\gamma^+]}(x,\xi)= \bar u(p_2 )\, {\cal I}^{[\gamma^+]}(x,\xi) \, u(p_1)\, ,
\end{eqnarray}
where
\begin{eqnarray}
\label{StInt}
{\cal I}^{[\gamma^+]}(x,\xi)\stackrel{{\rm def}}{=}
\int d\mu({\bf k}_T) \,\int dk^- \, \frac{\phi^+(k,\Delta)}
{D_1\, D_2\, D_3} \biggr|_{k^+=xP^+}\,
\end{eqnarray}
with
\begin{eqnarray}
&&\phi^+= \gamma_\alpha (\slashed{k} + \slashed{\Delta}/2)\gamma^+ (\slashed{k} - \slashed{\Delta}/2)\gamma_\alpha
\approx
- {\bf k}^2_T \gamma^+,
\nonumber\\
&& D_{1,3}=2k^- P^+ (x\mp\xi) - {\bf k}^2_T +i\epsilon ,
\nonumber\\
&&D_2=2k^- P^+ (x-1) - {\bf k}^2_T +i\epsilon \, .
\end{eqnarray}
We introduced the effective integration measure $d\mu({\bf k}_T)$ in Eq. (\ref{StInt}) in order to ensure the convergence of the corresponding integration.
Let us emphasize that this modification of the measure will not affect the results of our study. Indeed, our reasoning is also valid for the GPDs in the toy scalar model, considered, e.g., in Refs. \cite{Rad97, Rad-Sing}, because the numerator $\phi^+(k,\Delta)$ contains only ${\bf k}^2_T$ in the collinear kinematics.

Let us first carry out the integration over $k^-$ in (\ref{StInt}) in the complex plane. To this end,
we will analyze the analytical properties on the integrand, namely, the position of the poles in the complex plane of the variable $k^-$. We have (cf. \cite{Brod})
\begin{eqnarray}
\label{pol1}
&&k^-_1=- \frac{{\bf k}^2_T}{2P^+(\xi-x)}+i\epsilon,\,
k^-_2=- \frac{{\bf k}^2_T}{2P^+(1-x)}+i\epsilon,
\nonumber\\
&&k^-_3=- \frac{{\bf k}^2_T}{2P^+(\xi+x)}-i\epsilon
\end{eqnarray}
for $0<x<\xi$;
and
\begin{eqnarray}
\label{pol2}
&&k^-_2=- \frac{{\bf k}^2_T}{2P^+(1-x)}+i\epsilon,
\\
&&k^-_1= \frac{{\bf k}^2_T}{2P^+(x-\xi)}-i\epsilon,\,
k^-_3= \frac{{\bf k}^2_T}{2P^+(x+\xi)}-i\epsilon
\nonumber
\end{eqnarray}
for $x>\xi>0$.
For the negative fraction $x$, especially for the interval $-\xi<x<0$, the poles are situated similarly to the case of $0<x<\xi$; while for the interval $x<-\xi$ all poles
lie in the same semi-plane and, therefore, this region of the fraction does not contribute.
In (\ref{pol1}) and (\ref{pol2}), $k^-_{1,3}$ correspond to the quark poles while $k^-_2$---to
the gluon pole.

Thus, integrating over $k^-$ in its complex plane, we obtain
\begin{eqnarray}
\label{Phi-fin}
{\cal I}^{[\gamma^+]}(x,\xi)=\gamma^+\, \int (d{\bf k}^2_T)
\frac{\Psi^2({\bf k}^2_T)}{{\bf k}^2_T+\Lambda^2} \, H(x,\xi)\, ,
\end{eqnarray}
where
\begin{eqnarray}
\label{H}
&&H(x,\xi)=\theta(-\xi<x<\xi)\, \biggl[ \frac{\xi-x}{2\xi(1-\xi)}-\frac{1-x}{1-\xi^2}\biggr] -
\nonumber\\
&&\theta(\xi<x<1)\, \frac{1-x}{1-\xi^2}\, .
\end{eqnarray}
We here introduce the effective UV- and IR-regularizations following Ref. \cite{Rad-Sing}.
Eqs. (\ref{H}) can be split into the contributions of the quark and gluon poles separately:
\begin{eqnarray}
\label{SplitH}
&&H^{[...]}(x,\xi)=-\theta(-\xi<x<1)\frac{1-x}{1-\xi^2}\,,
\nonumber\\
&&H^{\{...\}}(x,\xi)=\theta(-\xi<x<\xi)\, \frac{\xi-x}{2\xi(1-\xi)}\, ,
\end{eqnarray}
where the ``anti-commutator part'' of the GPDs, $H^{\{...\}}(x,\xi)$, is related to the quark pole
contributions and the ``commutator part'',
$H^{[...]}(x,\xi)$,---to the gluon pole contribution. Indeed, consider the commutator contribution written in the following form
(see, (\ref{GPDs2})-(\ref{GPDs4}))
\begin{eqnarray}
\label{Hcom2}
&&\hspace{-0.5cm}H^{[...]}(x,\xi)= \int d^4k \, \delta(x-k \cdot n) \, {\cal A}^{[...]}(k)\, ,
\\
&&\hspace{-0.5cm}{\cal A}^{[...]}(k)=\hspace{-0.1cm}\frac{1}{2}\hspace{-0.1cm}
\int\hspace{-0.1cm} d^4z e^{i(k-\Delta/2).z}
\langle p_2|
[\bar\psi(0)\gamma^+, \psi(z)] |p_1 \rangle^H_C .
\nonumber
\end{eqnarray}
As before, we identify the initial and final states in Eq. (\ref{Hcom2}) with the electrons/quarks. Hence, we insert in Eq. (\ref{Hcom2}) the full set of the intermediate states
$ \sum_X |P_X \rangle^H \,{~}^H\langle P_X| = 1$ and obtain
\begin{eqnarray}
&&
{\cal A}^{[...]}(k)=\frac{1}{2} \sum_X\hspace{-0.5cm}\int\delta^{(4)}(k-P+P_X)\times
\nonumber\\
&&
\langle p_2|
\bar\psi(0)\gamma^+ |P_X \rangle^H\langle P_X|  \psi(0) |p_1 \rangle^H_C \, .
\end{eqnarray}
In order to be able to make use of perturbation theory, we transform to the interaction representation and keep the terms up to the $g^2$-order:
\begin{eqnarray}
&&{\cal A}^{[...]}(k)= \frac{1}{2}\sum_X\hspace{-0.5cm}\int\,\, \delta^{(4)}(k-P+P_X)
\nonumber\\
&&\langle p_2|
\mathrm{T}(\bar\psi(0)\gamma^+ \mathbb{S})|P_X \rangle \,
\langle P_X| \mathrm{T}( \psi(0) \mathbb{S})|p_1 \rangle_C
\stackrel{g^2\,{\rm PT}}{\Longrightarrow}
\nonumber\\
&& \delta((P-k)^2) \,
\bar u(p_2)\, \gamma\cdot\varepsilon^* S(k+\Delta/2)\, \gamma^+\times
\nonumber\\
&& S(k-\Delta/2)\, \gamma\cdot\varepsilon\, u(p_1)\, ,
\end{eqnarray}
where we have used the one-particle states $ |p_1 \rangle = b^+(p_1) |0 \rangle$ and $\langle p_2| = \langle 0| b^-(p_2)$, and we have chosen the one-boson (photon/gluon) state
as the intermediate state. Therefore, we obtain
\begin{eqnarray}
&&H^{[...]}(x,\xi)=\frac{1}{2} \int\, d{\bf k}^2_T \, dk^-
\delta(2k^-P^+(x-1)-{\bf k}^2_T) \,
\nonumber\\
&&\bar u(p_2)\, \gamma\cdot\varepsilon^* \, S(k+\Delta/2)\, \gamma^+\,
S(k-\Delta/2)\,
\gamma\cdot\varepsilon\, u(p_1)\,,
\nonumber
\end{eqnarray}
where we assume that $k^+=xP^+$. This expression can be re-written in the Heisenberg representation:
\begin{eqnarray}
&&H^{[...]}(x,\xi)=\frac{1}{2} \int\, d{\bf k}^2_T \, dk^-
\delta(2k^-P^+(x-1)-{\bf k}^2_T) \,
\nonumber\\
&&\langle p_2|
\bar\psi(0)\gamma^+ |P-k \rangle^H \,\langle P-k|  \psi(0) |p_1 \rangle^H_C
\,.
\end{eqnarray}
One can easily see that this expression is nothing else but the cut of the amplitude (\ref{AmpStInt}) in the photon/gluon propagator. To say the same thing in a different way,
this contribution comes from the diagrams where the photon/gluon propagator is replaced by its imaginary part (that is to say, it yields the gluon pole contribution).
In the same way we can show that the anti-commutator contribution is given by the quark pole contribution, or by picking up the cut in the quark propagator with the momentum $k+\Delta/2$.

\section{Polynomiality and Positivity for GPDs}

We are now in a position to address the polynomiality condition for (\ref{H}).
Calculating the corresponding moments of (\ref{H}), we have
\begin{eqnarray}
\label{PolCon1}
&&\int\limits_{-1}^{1} dx \, x^{2n} \, H(x,\xi) =
- \frac{2(1-\xi^{2n+2})}{(2n+1)(2n+2)(1-\xi^2)}=
\nonumber\\
&&c_0+c_2\xi^2 + ... + c_{2n}\xi^{2n}\,,
\nonumber\\
&&\int\limits_{-1}^{1} dx \, x^{2n+1} \, H(x,\xi) =
- \frac{2(1-\xi^{2n+2})}{(2n+2)(2n+3)(1-\xi^2)}
\nonumber\\
&&=d_0+d_2\xi^2 + ... + d_{2n}\xi^{2n}\, .
\end{eqnarray}
Let us stress that the box diagram itself cannot ensure the so-called $D$-term contribution which describes the resonance exchange diagram (see, e.g., Refs. \cite{PW, Rad-Sing}).
We will therefore treat, for a moment, the polynomiality of the GPDs as the expression of the corresponding moments
through the finite series with only even orders of $\xi$, see (\ref{PolCon1}).
By making use of the splitting (\ref{SplitH}), we can verify the polynomiality for each of the commutator and anti-commutator contributions. We have the following:
\begin{eqnarray}
\label{Hcom}
&&
\int\limits_{-1}^{1} dx x^{n}H^{[...]}(x,\xi) =
\frac{c_{-1}}{1-\xi}+\sum\limits_{k=0}^n a_{k}\xi^{k}\,,
\\
\label{Hanticom}
&&\int\limits_{-1}^{1} dx x^{n}H^{\{...\}}(x,\xi) =
-\frac{c_{-1}}{1-\xi}+\sum\limits_{k=0}^n b_{k}\xi^{k}\,
\end{eqnarray}
where $a_{2k-1}=-b_{2k-1}$.
One can see that neither the commutator contribution nor the anti-commutator contribution obeys the polynomiality separately.
In other words, we have the polynomiality only after summation of these two terms. We conclude, therefore, that
the anti-commutator contribution is necessary to satisfy the model independent polynomiality condition and, therefore, cannot be discarded by default. This is our principal result.

Now let us present the scheme how the new contribution arising from the $H^{\{...\}}$-term, Eq. (\ref{SplitH}), affects the positivity constraint, Ref. \cite{PST-Positiv}.
The full and comprehensive analysis will be presented in the forthcoming paper \cite{ACT-Positiv}.
The structure of the photon/gluon and the quark pole contributions in the factorized box diagram amplitude, where
the soft part has been calculated perturbatively, helps us to write down the Cauchy-Bunyakovsky-Schwarz inequality. We have
\begin{eqnarray}
\label{CBS-1}
&&\int d^4k \delta(x-k\cdot n) \delta((P-k)^2) \times
\nonumber\\
&&\biggl|
\lambda\langle P-k| \psi_+(0) |p_2 \rangle^H \hspace{-0.2cm}+
\langle P-k| \psi_+(0) |p_1 \rangle^H
\biggr|^2 +
\nonumber\\
&&\int d^4k \delta(x-k\cdot n) \delta((k+\Delta/2)^2)
\nonumber\\
&&
\biggl|
\lambda\langle k+\frac{\Delta}{2},p_1| \psi^\dag_+(0) |p_2 \rangle^H \hspace{-0.2cm}+
\langle k+\frac{\Delta}{2}| \psi^\dag_+(0) |0 \rangle^H
\biggr|^2\hspace{-0.2cm} \geq 0 .
\nonumber
\end{eqnarray}
Here, the light-cone components of the fermion fields are given by
$\psi_{\pm}=1/2\gamma^{\mp}\gamma^{\pm}\psi.$
The characteristic equation of Eq. (\ref{CBS-1}) takes the following form:
$\lambda^2 A + \lambda B + C \geq 0$,
where (using the crossing where needed)
\begin{eqnarray}
\label{A}
&&A=\int d^4k \delta(x-k\cdot n) \delta((P-k)^2) \times
\nonumber\\
&&\langle p_2|\psi^\dag_+(0) |P-k \rangle\langle P-k| \psi_+(0) |p_2 \rangle^H +
\\
&&\int d^4k \delta(x-k\cdot n) \delta((k+\Delta/2)^2)\times
\nonumber\\
&&\langle p_2,-p_1|
\psi_+(0) |k+\frac{\Delta}{2} \rangle
\langle k+\frac{\Delta}{2}| \psi^\dag_+(0) |-p_1,p_2 \rangle^H \, ,
\nonumber
\end{eqnarray}
\begin{eqnarray}
\label{B}
&&B=\int d^4k \delta(x-k\cdot n) \delta((P-k)^2) \times
\nonumber\\
&&
\langle p_2|
\psi^\dag_+(0) |P-k \rangle\langle P-k|  \psi_+(0) |p_1 \rangle^H +
\nonumber\\
&&\int d^4k \delta(x-k\cdot n) \delta((k+\Delta/2)^2)\times
\nonumber\\
&&
\langle p_2,-p_1|
\psi_+(0) |k+\frac{\Delta}{2} \rangle\langle k+\frac{\Delta}{2} |  \psi^\dag_+(0) |0 \rangle^H
\nonumber\\
\hspace{-0.5cm}&&+(p_1\leftrightarrow p_2),
\end{eqnarray}
and
\begin{eqnarray}
\label{C}
&&C=\int d^4k \delta(x-k\cdot n) \delta((P-k)^2) \times
\nonumber\\
&&
\langle p_1|
\psi^\dag_+(0) |P-k \rangle\langle P-k|  \psi_+(0) |p_1 \rangle^H +
\nonumber\\
&&\int d^4k \delta(x-k\cdot n) \delta((k+\Delta/2)^2)\times
\nonumber\\
&&
\langle 0|
\psi_+(0) | k+\frac{\Delta}{2} \rangle\langle k+\frac{\Delta}{2} |  \psi^\dag_+(0) |0 \rangle^H \, .
\end{eqnarray}
We now see that the first and the second terms of Eq. (\ref{B}) produce the ``commutator'' and ``anti-commutator'' GPDs, respectively, while the first and the second terms of Eqs. (\ref{A}) and (\ref{C}) correspond to the forward distributions and the vacuum expectations (the quark condensate). To satisfy the above-mentioned characteristic equation
we have to demand that $D=B^2-4AC\leq 0$, which is equivalent to the following inequality
(the corresponding normalization of $q(x)$ is implied):
\begin{eqnarray}
\label{CBS-3}
&&\biggl[ H^{[...]}_{S(A)}(x,\xi) + H^{\{...\}}_{S(A)}(x,\xi)
\biggr]^2 \leq
\nonumber\\
&&\biggl[ q(x_2) + D(x_2)
\biggr]\, \biggl[ q(x_1) + C(x_1)
\biggr] \, ,
\end{eqnarray}
where we introduced the symmetrized and anti-symmetrized in $x\leftrightarrow -x$ combinations of the corresponding GPDs
and performed the re-scaling of the fractions, see \cite{PST-Positiv}. After the summation over the intermediate states,  the functions $D(x)$ and $C(x)$, Eq. (\ref{CBS-3}),
take the following form
\begin{eqnarray}
&&D(x)= \int d^4k \, \delta(x-k.n)\, d^4z \ e^{i(k-\Delta/2) \cdot z}\times
\nonumber\\
&&\langle p_2,p_1 |
\psi_{+}(z)\psi^{\dag}_{+}(0) |p_2,p_1 \rangle^H\,,
\nonumber\\
&&C(x)=\int d^4k \, \delta(x-k.n)\, d^4z \ e^{i(k-\Delta/2) \cdot z} \times
\nonumber\\
&&
\langle 0 |
\psi_{+}(z) \psi^{\dag}_{+}(0) |0 \rangle^H\, .
\end{eqnarray}

\section{Conclusions}

To conclude, we have found that in the collinear kinematics and in the factorization
regime with $t \approx 0$,
the matrix element of the fermion anti-commutator does not vanish. We have demonstrated, moreover, that the existence of this contribution is dictated by the polynomiality condition for the GPDs.
Furthermore, we have obtained a new possible constraint for the GPDs
wherein the new contributions from the forward distribution and the quark condensate are included.

The authors thank A.V.~Efremov, L.~Szymanowski, O.V.~Teryaev and S.~Wallon for useful discussions and comments. This work is supported in part by the RFBR (grant 12-02-00613).

\end{document}